\documentclass[11pt,a4paper]{article}
\usepackage{amsmath,amssymb,amsfonts,a4wide,graphicx,bm,times}
\usepackage{mcite}

\makeatletter \let\old@startsection=\@startsection
\renewcommand{\@startsection}[6]
{\old@startsection{#1}{#2}{#3}{#4}{#5}{#6\mathversion{bold}}}
\makeatother

\makeatletter

\@addtoreset{equation}{section}
\makeatother
\let\refOld\ref
\renewcommand{\ref}[1]{(\refOld{#1})}

\newcommand{\superp}[2]{\genfrac{}{}{0pt}{}{#1}{#2}}

 \def\Disc{\text{Disc }}
 \def\d{\delta}

 \def\Im{{\rm Im ~}}
 \def\p{\partial}
 
 \def\a{\alpha}
 \def\b{\beta}
 \def\g{\gamma}
 \def\d{\delta}
 \def\e{\epsilon}

 \def\l{\lambda}

 \def\s{\sigma}
 \def\t{\tau}

 \def\G{\Gamma}

 \def\S{\Sigma}
 
 \def\O{\Omega}

\def\la{\left\langle}
\def\ra{\right\rangle}

\def\laN{\left\langle N\left|}
\def\raN{\right|N\right\rangle}
\def\hf{\dfrac{1}{2}}
\def\Op{\mathcal{O}}
\def\CD{\mathcal{D}}

\def\implies{\quad\Rightarrow\quad}

\def\CF{\mathcal{F}}

\def\CZ{{\mathcal{Z}}}

\def\bens{$\b$-ensemble}

\def\brho{\bar\rho}

\def\CS{\mathcal{S}}

\def\bq{\bar{q}}
\def\SC{\CS_\text{C}}
\def\SGC{\CS_\text{GC}}

\def\ZGC{\CZ_\text{GC}}
\def\ZC{\CZ_\text{C}}
\def\FGC{\CF_\text{GC}}
\def\FC{\CF_\text{C}}
\def\bY{\bar Y}
\def\Aut{\text{Aut}}

\begin{document}
\begin{titlepage}
\renewcommand{\thefootnote}{\fnsymbol{footnote}}

\vspace*{-2cm}
\begin{flushright}
APCTP Pre2013-017
\end{flushright}

\vspace*{1cm}
    \begin{Large}
       \begin{center}
         {Notes on Mayer Expansions and Matrix Models}
       \end{center}
    \end{Large}
\vspace{0.7cm}

\begin{center}
Jean-Emile B{\sc ourgine}\footnote
            {
e-mail address : 
jebourgine@apctp.org}\\
      
\vspace{0.7cm}                    
{\it Asia Pacific Center for Theoretical Physics (APCTP)
}\\
{\it Pohang, Gyeongbuk 790-784, Republic of Korea}
\end{center}

\vspace{0.7cm}

\begin{abstract}
\noindent
Mayer cluster expansion is an important tool in statistical physics to evaluate grand canonical partition functions. It has recently been applied to the Nekrasov instanton partition function of $\mathcal{N}=2$ 4d gauge theories. The associated canonical model involves coupled integrations that take the form of a generalized matrix model. It can be studied with the standard techniques of matrix models, in particular collective field theory and loop equations. In the first part of these notes, we explain how the results of collective field theory can be derived from the cluster expansion. The equalities between free energies at first orders is explained by the discrete Laplace transform relating canonical and grand canonical models. In a second part, we study the canonical loop equations and associate them to similar relations on the grand canonical side. It leads to relate the multi-point densities, fundamental objects of the matrix model, to the generating functions of multi-rooted clusters. Finally, a method is proposed to derive loop equations directly on the grand canonical model.
\end{abstract}
\vfill

\end{titlepage}
\vfil\eject

\setcounter{footnote}{0}

\section{Introduction}
The AGT correspondence \cite{Alday2009} implies a relation between the canonical partition function of a \bens\ and the grand canonical partition function of a generalized matrix model. The former represents a correlator of Liouville theory, according to the proposal of Dijkgraaf and Vafa \cite{Dijkgraaf2009}, further investigated in \cite{Fujita2009,Mironov2010c,Mironov2010e,Itoyama2010a,Itoyama2011,Nishinaka2011,Bonelli2011,Bonelli2011a,Baek2013}. The latter describes the instanton partition function of a 4d $\mathcal{N}=2$ supersymmetric gauge theory in the $\Omega$-background, as derived using localization techniques in \cite{Nekrasov2003}. Here the term 'generalized matrix model' do not pertain to a matrix origin for the model, but instead refers to a set of models that can be studied using techniques initially developed in the realm of matrix models. Among these techniques, the \textit{topological recursion} \cite{Eynard2007a} exploits the invariance of the integration measure to derive a tower of nested equations satisfied by the correlators of the model. These equations, referred as \textit{loop equations}, are solved employing methods from algebraic geometry. This technique has recently been extended to a wide spectrum of coupled integrals models in \cite{Borot2013}.

In a suitable limit of the \bens, AGT-equivalent to the Nekrasov-Shatashvili (NS) limit of the $\Omega$-background \cite{Nekrasov2009}, loop equations are no longer algebraic but first order linear differential equations.\footnote{Except for the first (planar) equation, which is a Ricatti equation, therefore non-linear. It is equivalent to a Schr\"odinger equation, i.e. a linear differential equation of second order.} In this context, the \bens\ is a natural quantization of the Hermitian matrix model, to which it reduces at $\b=1$. The first element of this tower of differential equations has been mapped to the TQ relation derived in \cite{Poghossian2010,Fucito2011,Fucito2012} that describes the dual SUSY gauge theory in the NS limit \cite{Zenkevich2011,Mironov2012a,Mironov2012b,Bourgine2012a}. It is then natural to ask about the existence of a structure similar to loop equations on the gauge side of the correspondence.\footnote{Such a structure should be related to the invariance of Nekrasov partition functions under transformations representing the SHc algebra uncovered in \cite{Kanno2013} (see also \cite{Kanno2011,Kanno2012}).} But so far, the loop equation technique has not been applied to grand canonical matrix models. On the other hand, the cluster expansion of Mayer and Montroll \cite{Mayer1941} has been successfully employed to derived an effective action relevant to the NS limit \cite{Nekrasov2009}. Can we relate this cluster expansion to the topological expansion of a generalized matrix model? Is there an equivalent of the loop equations technique on the grand canonical side? And more generally, how do canonical and grand canonical coupled integrals relate to each other? These are the issues we propose to address in these notes.

For this purpose, we consider the following grand canonical generalized matrix model,
\begin{equation}\label{def_CZ}
\ZGC(\bq)=\sum_{N=0}^\infty{\dfrac{\bq^N}{N!}\ZC(N)},\quad \ZC(N)=\int_{\mathbb{R}^N}{\prod_{i=1}^NQ(\phi_i)\dfrac{d\phi_i}{2i\pi}\prod_{\superp{i,j=1}{i<j}}^NK(\phi_i-\phi_j)}.
\end{equation}
In analogy with the Nekrasov partition function, integrals are understood as contour integrals over the real line. The potential $Q(x)$ and the kernel $K(x)$ are free of singularities over the real axis.\footnote{In the case of real singularities, a prescription should be given to move away the poles from the contour by a small imaginary shift.} We propose to study the expansion of $\ZGC(q)$ when the kernel is close to one. More precisely, we assume the form
\begin{equation}\label{dev_K}
K(x)=1+\e f(x),\quad \e\to 0,
\end{equation}
with $f$ an even function, non-vanishing at $x=0$. Although the results of these notes are very general, what we have in mind for the function $f$ is typically
\begin{equation}
f(x)=\dfrac1{x^2-\g^2},\quad \Im\g\neq0.
\end{equation}
It is crucial for our considerations that $f$ is independent of $\e$. In this way, we exclude a class of models more relevant to the study of Nekrasov partition functions. For instance, setting $\e=\g^2$, one recovers the model proposed by J. Hoppe in \cite{Hoppe1982}. This model is a one-parameter version of the Nekrasov partition function that depends on two $\O$-background equivariant deformation parameters $\e_1$ and $\e_2$ \cite{Kazakov1998,Hoppe1999}. As $\e\to0$, it exhibits a phenomenon referred as \textit{instanton clustering} in the context of SUSY gauge theories \cite{Nekrasov2009}. It corresponds to poles coming from the kernel and pinching the integration contour. Such poles should be avoided by a deformation of the contour, picking up the corresponding residues. As a result, terms of the $\e$-expansions we are considering are reshuffled and the results presented here are no longer valid.

These notes are organized as follows. In the second section, we compare the Mayer cluster expansion of the grand canonical model with the collective field theory describing the large $N$ limit of the canonical model. Taking the coupled limit $\e\to0$ and $N\to\infty$ with $N\e$ fixed, we derive relations between the free energies at first orders. These relations are a consequence of the fact that the grand canonical partition function is the discrete Laplace transform of the canonical one. We go on with the study of the canonical loop equations. We show that they relate to graphical identities between generating functions of rooted clusters. Such generating functions show up in the Mayer expansion and are identified with the multi-point densities. Finally, we present a technique to derive directly the grand canonical loop equations. The main results are summarized in the concluding section.

\section{Comparison of the free energies at first orders}
\subsection{Mayer expansion of the grand canonical model}
\begin{figure}[!t]
\centering
\includegraphics[width=10cm]{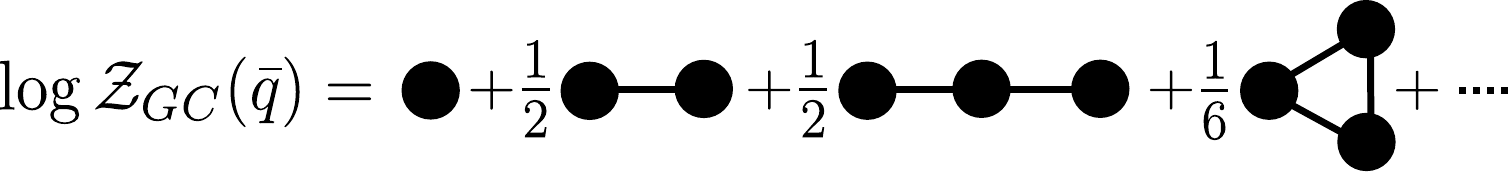}
\caption{First orders in the Mayer expansion of the free energy.}
\label{ZGC}
\end{figure}

The cluster expansion was introduced by Mayer and Montroll as a way to compute the free energy knowing the form of the interaction between particles \cite{Mayer1941} (see also the book \cite{Mayer1940} and the excellent review by Andersen \cite{Andersen1977}). It allows to derive the equation of state for various types of fluids. To do so, the kernel is expanded in $\e$, which corresponds to strength of molecular interactions in the case of non-ideal gases. The terms of the series consist of coupled integrals with the kernel $f$ instead of $K$, and their expression is encoded into clusters. Here, a cluster is a set of vertices connected by at most one link. The partition function is a sum over disconnected clusters, but after taking the logarithm the summation is restricted to connected ones. We denote by $C_l$ a generic connected cluster with $l$ vertices,  $E(C_l)$ the set of its links (or edges) and $V(C_l)$ the set of vertices. To each vertex $i$ of a cluster is associated an integration over the particle of coordinate $\phi_i$ with measure $\bq Q(\phi_i)d\phi_i/2i\pi$. The edge $<ij>$ between particles $i$ and $j$ represents the kernel $\e f(\phi_i-\phi_j)$. Thus, the logarithm of the partition function writes
\begin{equation}\label{sum_clusters}
\log\ZGC(\bq)=\sum_{l=0}^\infty{\bq^l\sum_{C_l}\dfrac1{\s(C_l)}\int{\prod_{i\in V(C_l)}Q(\phi_i)\dfrac{d\phi_i}{2i\pi}\prod_{<ij>\in E(C_l)}\e f(\phi_i-\phi_j)}},
\end{equation}
where the symmetry factor $\s(C_l)$ is the cardinal of the group of automorphism for the cluster, i.e. the number of permutations of vertices that leave $C_l$ invariant. The first terms of the expansion and their symmetry coefficients are given in figure \refOld{ZGC}.

The Mayer expansion \ref{sum_clusters} is an expansion at small (bare) fugacity $\bq$. We would like to reformulate it as a $\bq$-exact expansion in the parameter $\e$. We will also renormalize the fugacity, keeping $q=\bq \e$ fixed. By analogy, $\bq$ would encode the gauge coupling constant of the Nekrasov partition function, and the Mayer cluster expansion is an expansion upon the number of instantons. More precisely, $\bq$ would correspond to $q_\text{gauge}(\e_1+\e_2)/\e_1\e_2$ and should be renormalized by a factor $\e_2$ in the NS limit $\e_2\to0$. In this context, the $\e$-expansion we study corresponds to an expansion in the $\Omega$-background parameter $\e_2$, or in the AGT dual, to the semi-classical expansion of Liouville correlators.

\begin{figure}[!t]
\centering
\includegraphics[width=9cm]{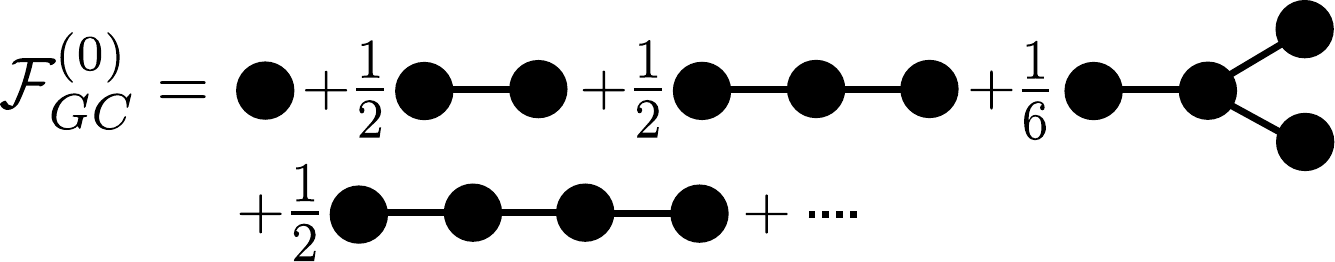}
\caption{The leading order free energy as a sum over trees.}
\label{ZGC_trees}
\end{figure}

Since each link brings a factor of $\e$, at first order only the clusters with a minimal number of links contribute. These clusters, denoted $T_l$, have a tree structure, with $l-1$ links for $l$ vertices. Thus, at first order in $\e$ the free energy is given by the following sum over trees,
\begin{equation}\label{sum_trees}
\FGC^{(0)}(q)=\lim_{\e\to0}\e\log\ZGC(\bq)=\sum_{l=0}^\infty{q^l\sum_{T_l}\dfrac1{\s(T_l)}\int{\prod_{i\in V(T_l)}Q(\phi_i)\dfrac{d\phi_i}{2i\pi}\prod_{<ij>\in E(T_l)}f(\phi_{ij})}},
\end{equation}
where we used the shortcut notation $\phi_{ij}=\phi_i-\phi_j$. Note that we have renormalized the free energy by a factor of $\e$, which is reminiscent of the volume of the $\Omega$-background $\e_1\e_2$ by which the prepotential should be multiplied in order to be finite in the $\mathbb{R}^4$ limit $\e_1,\e_2\to0$. The first terms of this expansion are given in the figure \refOld{ZGC_trees}.

\begin{figure}[!t]
\centering
\includegraphics[width=9cm]{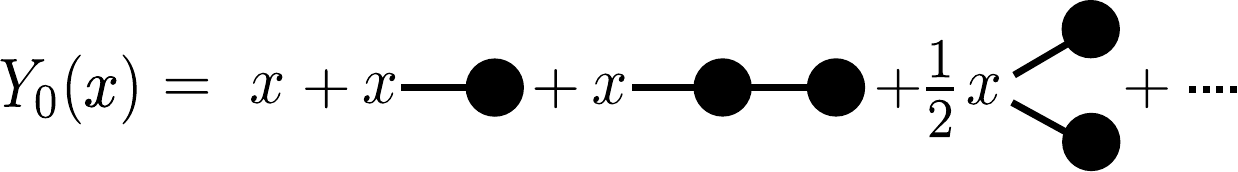}
\caption{First terms in the expansion of the rooted tree generating function $Y_0(x)$.}
\label{Y0_def}
\end{figure}

To evaluate $\FGC^{(0)}$, it is convenient to consider the generating function of rooted trees $T_l^x$, defined as
\begin{equation}\label{def_Y0}
Y_0(x)=q Q(x)\sum_{l=0}^\infty\sum_{T_l^x}\dfrac1{\s(T_l^x)}\int{\prod_{i\in V(T_l^x\smallsetminus\{x\})}q Q(\phi_i)\dfrac{d\phi_i}{2i\pi}\prod_{<ij>\in E(T_l^x\smallsetminus\{x\})}f(\phi_{ij})\prod_{<xi>\in E(T_l^x)}f(x-\phi_i)},
\end{equation}
where with a slight abuse of notations we denoted the root and its coordinate by the same letter $x$. The first order terms of this expansion are given in figure \refOld{Y0_def}. This function is interpreted as a tree-level dressed vertex. We should also emphasize that 'rooting' a tree, or marking a vertex, reduces the symmetry factor $\s(T_l^x)\leq\s(T_l)$ since automorphisms are now constraint to leave the root, or the marked vertex, invariant.

\begin{figure}[!t]
\centering
\includegraphics[width=5cm]{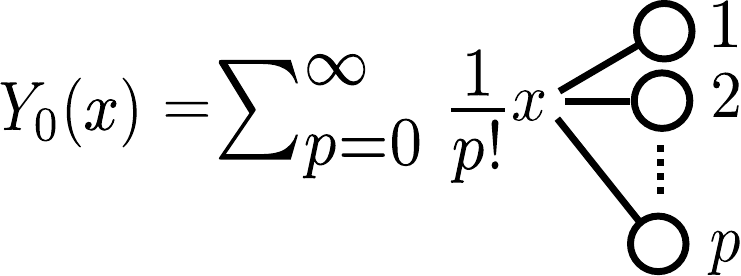}
\caption{Graphical representation of the recursion relation obeyed by $Y_0$.}
\label{Y0_rec}
\end{figure}

The function $Y_0(x)$ obeys an integral equation that can be obtained as follows. Let us assume that the root $x$ is directly connected to $p$ vertices, and sum over the possible numbers $p$. Each of these $p$ vertices is the root of a new tree, and we deduce the relation,
\begin{equation}
Y_0(x)=q Q(x)\sum_{p=0}^\infty{\dfrac1{p!}\left(\int_\mathbb{R}{\dfrac{dy}{2i\pi}f(x-y)Y_0(y)}\right)^p},
\end{equation}
graphically represented on figure \refOld{Y0_rec}. The symmetry factor $p!$ takes into account the possibility of permuting the $p$ vertices. Performing the summation, and taking the logarithm, we obtain the integral equation satisfied by $Y_0$,
\begin{equation}\label{int_equ}
\log\left(\dfrac{Y_0(x)}{q Q(x)}\right)=\int_\mathbb{R}{\dfrac{dy}{2i\pi}f(x-y)Y_0(y)}.
\end{equation}

It remains to relate the free energy to the generating function $Y_0$. This is done using the following formula due to B. Basso, A. Sever and P. Vieira \cite{Basso2013},\footnote{This very useful formula was brought to my knowledge by B. Basso. So far, we were unable to find a proper reference in the previous literature. However, similar considerations were presented in \cite{Drouffe1983}.}
\begin{equation}\label{Basso}
\FGC^{(0)}(q)=\int_\mathbb{R}{Y_0(x)\dfrac{dx}{2i\pi}-\hf\int_{\mathbb{R}^2}}{\dfrac{dx}{2i\pi}\dfrac{dy}{2i\pi}Y_0(x)Y_0(y)f(x-y)}.
\end{equation}
It is easy to see that both terms in the RHS will produce a sum over clusters weighted by the same integrals as in \ref{sum_trees}, but with different symmetry factors. A combinatorial proof of this formula is given in appendix \refOld{App_Basso}.

It is useful to reformulate the previous expression \ref{Basso} of the free energy at first order as the value of an effective action $\SGC[Y_0]$ at its extremum $Y_0^\ast$,
\begin{equation}
\FGC^{(0)}(q)=\SGC[Y_0^\ast],\quad \text{such that}\quad \left.\dfrac{\d \SGC}{\d Y_0}\right|_{Y_0=Y_0^\ast}=0,
\end{equation}
This action is obtained after introducing the integral equation \ref{int_equ} into \ref{Basso},
\begin{equation}\label{eff_action}
\SGC[Y_0]=\hf\int_{\mathbb{R}^2}{\dfrac{dx}{2i\pi}\dfrac{dy}{2i\pi}Y_0(x)Y_0(y)f(x-y)}-\int_\mathbb{R}{\dfrac{dx}{2i\pi}Y_0(x)\left[\log Y_0(x)-1\right]}+\int_\mathbb{R}{\dfrac{dx}{2i\pi}Y_0(x)\log\left(q Q(x)\right)}.
\end{equation}
It is remarkable that the saddle point equation derived from this action is nothing else than the integral equation \ref{int_equ}. It is also worth noticing that when instanton clustering phenomenon is taken into account, one arrive at a similar expression, with logarithms replaced by dilogarithms in the second term. For instance, the effective action derived by Nekrasov and Shatashvili to describe $\mathcal{N}=2$ SYM reads
\begin{equation}\label{action_NS}
\mathcal{S}_\text{NS}[\rho]=\hf\int{\rho(x)\rho(y)G(x-y)}+\int{\left[\text{Li}_2(1-e^{-\rho(x)})-\rho(x)\log(1-e^{-\rho(x)})\right]dx}+\int{\rho(x)\log\left(qQ(x)\right)dx},
\end{equation}
where we used the notations of \cite{Nekrasov2009}. Expanding the middle term in $\rho$, we recover at the order $o(\rho^2)$ the action \ref{eff_action} obtained previously. This type of 'cut-off' term for the action has been studied in \cite{Bourgine2012a}.

\paragraph{Subleading order} We now focus on $\FGC^{(1)}$, the second order term in the $\e$-expansion of the free energy at fixed $q$,
\begin{equation}
\e\log\ZGC(\bq)=\sum_{k=0}^\infty{\e^{k}\FGC^{(k)}(q)}.
\end{equation}
At this order, clusters that contribute have $l$ links for $l$ vertices, which means that they have exactly one cycle. Such clusters will be denoted $S_l$. The relevant terms of the Mayer expansion for the free energy are
\begin{equation}
\FGC^{(1)}(q)=\sum_{l=3}^\infty{q^l\sum_{S_l}\dfrac1{\s(S_l)}\int{\prod_{i\in V(S_l)}Q(\phi_i)\dfrac{d\phi_i}{2i\pi}\prod_{<ij>\in E(S_l)}f(\phi_{ij})}}.
\end{equation}
The $q$-expansion starts here at $l=3$ since at least three vertices are needed to form a cycle. As we go to higher orders in $\e$, more vertices will be needed to form the cycles, leading to a higher first order term in $q$. Thus, $\FGC^{(0)}$ fully determines the $q$-expansion of the free energy up to order $O(q^2)$, and $\FGC^{(0)}+\e \FGC^{(1)}$ up to $O(q^3)$.

\begin{figure}[!t]
\centering
\includegraphics[width=5cm]{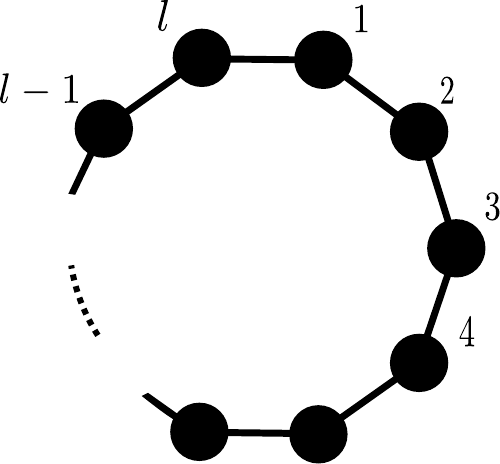}
\caption{Pure cycle clusters.}
\label{decagone}
\end{figure}

To evaluate the summation over clusters $S_l$, we first consider the clusters depicted on figure \refOld{decagone}, and for which all vertices belong to the cycle. Such clusters have a symmetry factor of $\s(S_l)=2l$ due to the invariance under $l$ rotations, and a reflexion symmetry. Their contribution writes
\begin{equation}\label{contrib_cycle}
\dfrac1{2l}\int{\prod_{i=1}^l f(\phi_i-\phi_{i+1}) qQ(\phi_i)\dfrac{d\phi_i}{2i\pi}},
\end{equation}
where indices are taken modulo $l$. 

All the clusters of type $S_l$ may be obtained by dressing the vertices of a pure cycle cluster by appropriate trees. Summing over the dressing possibilities boils down to replace $qQ(\phi)$ in the formula \ref{contrib_cycle} by the tree-level dressed vertex $Y_0(\phi)$. The expression for the free energy correction follows,
\begin{equation}\label{F1_serie}
\FGC^{(1)}(q)=\sum_{l=3}^\infty{\dfrac1{2l}\int{\prod_{i=1}^l f(\phi_i-\phi_{i+1})Y_0(\phi_i)\dfrac{d\phi_i}{2i\pi}}}.
\end{equation}
This is actually the expansion of the logarithm of a Fredholm determinant where the first two terms are missing. Taking the exponential, we find
\begin{equation}\label{F1}
e^{\FGC^{(1)}(\bq)}=\dfrac{e^{-\frac12f(0)\int{\frac{dx}{2i\pi}Y_0(x)}}}{\sqrt{\det\left[\d(x-y)-\frac{1}{2i\pi} f(x-y)Y_0(y)\right]}}\exp\left(-\dfrac14\int{\dfrac{dx}{2i\pi}\dfrac{dy}{2i\pi}Y_0(x)Y_0(y)f(x-y)^2}\right).
\end{equation}
The two missing terms correspond to a tadpole (a vertex with a link looping back to it) and two vertices doubly connected.

\subsection{Collective field theory of the canonical model}
The action \ref{eff_action} obtained above describes a Dyson gas of particles with the non-singular interaction $f$ at $\b=0$ \cite{Dyson1962}. It is also the effective action of a collective field theory for a generalized matrix model at large $N$ \cite{Jevicki1981}. We will show here that the corresponding matrix model is simply the canonical model $\ZC$ defined in the introduction (up to minor corrections). The fact that grand canonical and canonical models share the same effective action further leads to relate the rooted vertex generating function at tree level $Y_0$ with the collective field at large $N$.

At first order in $\e$, the canonical partition function is equivalent to
\begin{equation}\label{ZC_order1}
\ZC(N,\e)\simeq\int{\prod_{i=1}^N Q(\phi_i)\dfrac{d\phi_i}{2i\pi}\prod_{\superp{i,j=1}{i<j}}^Ne^{\e f(\phi_{ij})}}.
\end{equation}
The collective field is by definition a generating function of invariants under the permutation of eigenvalues. It is convenient to use the eigenvalue density,
\begin{equation}\label{def_rho0}
\rho_0(x)=\dfrac1{N}\sum_{i=1}^N\d(x-\phi_i),
\end{equation}
that has been normalized to one. It is usual for matrix models to assume that in the large $N$ limit, eigenvalues condense into a finite union of connected sets, typically a union of intervals for Hermitian matrices. This set $\G$ is the support of a continuous eigenvalue density $\rho_0$ obtained as the large $N$ limit of the finite densities defined in \ref{def_rho0}. Depending on the explicit form of potential and interaction, this assumption might not be valid. We will nonetheless work in this framework, the results derived following this approach being consistent with those obtained on the grand canonical model.

In the collective field theory approach, the canonical free energy is given at first order by the extrema of an effective action $\SC$,
\begin{equation}\label{eom_SC}
e^{N}\ZC(N,\e)\simeq e^{N\SC[\rho_0^\ast]},\quad \p_x\left.\dfrac{\d \SC}{\d\rho_0(x)}\right|_{\rho_0=\rho_0^\ast}=0.
\end{equation}
The factor $e^{N}$ has been introduced here to facilitate later comparison with the previous subsection. The canonical action is a sum of three terms,
\begin{equation}\label{def_SMM}
\SC[\rho_0]=\hf\e N\int{f(x-y)\rho_0(x)\rho_0(y)dxdy}+\int{\rho_0(x)\log\left(\dfrac{Q(x)}{2i\pi}\right)dx}-\int{\rho_0(x)\left[\log\rho_0(x)-1\right]dx}.
\end{equation}
The derivation of the first two terms is rather straightforward since it is sufficient to write down the integrand of \ref{ZC_order1} in an exponential form, and replace the sum over eigenvalues by integrals of the density. The third term corresponds to the entropic term introduced by Dyson in \cite{Dyson1962}. It is a Gibbs factor, coming from the fact that the Coulomb gas charges are indistinguishable. Following \cite{Jevicki1981,Lechtenfeld1991}, it is re-derived in the appendix \refOld{App_entropic} as a Jacobian in the change of measure from the discrete set of variables $d\phi_i$ to the functional integral over $D[\rho_0]$. In the case of Hermitian matrix models, such entropic factors cancel with the energetic term coming from the regularization of the kernel at coinciding eigenvalues. However, here $f(x)$ is finite at $x=0$ and cancellation does not occur.\footnote{In the case of \bens, there is also no cancellation, and these terms are responsible for the presence of the resolvent derivative in the loop equation \cite{Bourgine2012}.}

Comparing \ref{def_SMM} and \ref{eff_action}, we deduce that the effective actions are equivalent, upon the identification of $Y_0(x)$ with the density $2\pi i\rho_0(x)$, and provided we set $\e=1/N$. However, by definition the density $\rho_0$ is normalized to one, and this identification would require $Y_0$ to have also a unite norm. To resolve this issue, we introduce the norm $\a$ of $Y_0$ and identify as follows,
\begin{equation}
Y_0(x)=2i\pi\a\rho_0(x),\quad\text{with}\quad \a=\int{Y_0(x)\dfrac{dx}{2i\pi}}.
\end{equation}
This identification requires to set $N\e=\a=O(1)$ in the limit $N\to\infty$ and $\e\to0$. This relation signifies that the summation \ref{def_CZ} defining the grand canonical model is dominated at $\e\to0$ by the term with $N=\a/\e$ variables. Similarly, the Nekrasov partition function expressed as a sum over Young tableaux is dominated in the Seiberg-Witten limit $\e_1,\e_2\to0$ by a partition with $N\sim 1/\e_1\e_2$ boxes \cite{Nekrasov2003a,Nekrasov2012}. It also justifies the approach of \cite{Poghossian2010,Fucito2011,Fucito2012,Ferrari2012a,Bourgine2012a} to the study of the NS limit.

Under the previous identification between the dressed vertex $Y_0$ and the density $\rho_0$, canonical and grand canonical actions are related through
\begin{equation}
\SC[\rho_0]=\dfrac1\a\SGC[Y_0]+\log(\a/q).
\end{equation}
The term proportional to $\log q$ is missing from the action \ref{def_SMM}, but it can be introduced by hand, exploiting the fact that the density $\rho_0$ is normalized to one. In this case, $\log q$ plays the role of a Lagrange multiplier imposing the unit norm.

The two actions $\SC$ and $\SGC$ produce equivalent equations of motions, and the free energies satisfy at first order
\begin{equation}\label{Legendre}
\FC^{(0)}+\log q=\dfrac1{\a}\FGC^{(0)}(q)+\log\a-1,\quad \text{with}\quad \FC^{(0)}=\lim_{N\to\infty}\dfrac1N\log \ZC(N,\e=\a/N).
\end{equation}
Since $\FC$ depends on $N$ but not on $q$, and the opposite for $\FGC$, this relation only holds for a specific value of $N(q)$ or $q(N)$. More comments on this will follow in the next subsection where this relation is re-derived by exploiting the fact that $\ZGC$ is the discrete Laplace transform of $\ZC$.

\paragraph{One loop determinant} The subleading, or genus one, correction to the free energy can also be computed in the framework of the collective field theory. There are two types of corrections. The first one corresponds to amend the canonical action by a subleading term $\d\SC$, and the second type to the Gaussian fluctuations around the saddle point. The modification of the action is due to an earlier kernel approximation that should now be refined. Indeed, at the second order in $\e$ the approximation \ref{ZC_order1} of $\ZC$ is no longer acceptable and must be replaced by
\begin{equation}\label{def_ZN_II}
\ZC(N,\e)\simeq e^{-\frac12N\e f(0)}\int{\prod_{i=1}^N Q(\phi_i)\dfrac{d\phi_i}{2i\pi}\prod_{i,j=1}^Ne^{\frac12\e f(\phi_{ij})-\frac14\e^2 f(\phi_{ij})^2}}.
\end{equation}
The correction to the kernel is responsible for an additional contribution to the canonical action,
\begin{equation}
\d \SC=-\dfrac14\e^2N\int{f(x-y)^2\rho_0(x)\rho_0(y)dxdy},
\end{equation}
which reproduces the term in the second exponential of the expression \ref{F1} for $\FGC^{(1)}(\bq)$, provided we set again $\e N=\a$. The first exponential corresponds to the factor in front of the integrals in \ref{def_ZN_II}, and comes from the diagonal part of the kernel.

It is well known that the integration of Gaussian fluctuations around the classical solution produces the inverse square root of (minus) the Hessian matrix determinant,
\begin{equation}
\det\left[-\dfrac{\d^2 \SC}{\d\rho_0(x)\d\rho_0(y)}\right]=e^{-\int{\log \rho_0(x)dx}}\det\left[\d(x-y)-\e N\rho_0(x)f(x-y)\right].
\end{equation}
The prefactor involving the integral of $\log \rho_0(x)$ cancels with the sub-leading order of the entropic term computed in appendix \refOld{App_entropic}, formula \ref{sub_lead}. The remaining determinant reproduces the one which appears in \ref{F1}, upon the identification $Y_0(x)=2i\pi\a\rho_0(x)$ and $\e N=\a$. Gathering all contributions, we find 
\begin{equation}\label{ZC_1}
e^{\FC^{(1)}}= e^{-\frac12\a f(0)}\dfrac{e^{-\frac14\a^2\int{f(x-y)^2\rho_0(x)\rho_0(y)dxdy}}}{\sqrt{\det\left[\d(x-y)-\a\rho_0(x)f(x-y)\right]}}.
\end{equation}
Comparing with \ref{F1}, we conclude that the sub-leading contributions to the free energy of both models are equal. Again, this equality holds only for a specific value of $q(N)$ or $N(q)$.

\subsection{Discrete Laplace transform at large $N$}\label{sec_GC_VS_C}
The observed relations between free energies at first orders originate in the discrete Laplace transform, also called $Z$-transform, performed in \ref{def_CZ} to define the grand canonical model. This transformation can be inverted by considering a contour integral over $q=\bq\e$ circling the origin,
\begin{equation}\label{Z_trans}
\ZC(N,\e)=N!\oint_0{\dfrac{dq}{2i\pi q}q^{-N}\e^{N}\ZGC(\bq,\e)}.
\end{equation}
In the large $N$ limit, it is possible to evaluate the integral using a saddle point technique \cite{Zia2009}, and the relation between grand canonical and canonical free energies is a simple Legendre transform,
\begin{equation}\label{Legendre_FE}
N\left(\FC(N,\e)+1-\log(N\e)\right)\simeq\e^{-1}\FGC(q,\e)-N\mu.
\end{equation}
On the LHS, the additional terms are due to the factor $1/N!$ and can be absorbed in the definition of $\FC(N,\e)$. Under this transformation, the number of particles $N$ and the chemical potential $\mu=\log q$ are conjugate variables. They are related through the saddle point equation,
\begin{equation}\label{saddle}
q\p_{q}\FGC(q,\e)=N\e.
\end{equation}
This equation can be solved in terms of $N(q)$ and \ref{Legendre_FE} provides the grand canonical free energy knowing the canonical one. Inverting the Legendre transform, $\FC$ can be derived from $\FGC$ with $\mu=-\p_N(N\FC-N\log N)$. 

In the previous considerations, $\e$ was a simple spectator. The novelty in these notes is to tune the parameter $\e$ toward zero as the number of particles is sent to infinity, keeping $\e N=\a$ fixed. The Legendre transformation \ref{Legendre_FE} survives this limit and produces the relation \ref{Legendre} between first orders free energies. In this limit, the conjugate variables are $\a$ and $\mu$. It is also interesting to note that the saddle point equation \ref{saddle} gives the normalization condition for the dressed vertex $Y(x)$,
\begin{equation}\label{qdq}
q\p_{q}\FGC(q,\e)=\int{Y(x)\dfrac{dx}{2i\pi}}=\a.
\end{equation}
The dressed vertex is the generating function of connected rooted clusters. Its expression is given by \ref{def_Y0} after replacing the summation over rooted tree by a general summation over rooted clusters $C_l^x$ with appropriate $\e$ factors. The first equality in \ref{qdq} is shown in the appendix \refOld{App_qdq} using the Mayer expansion \ref{sum_clusters} of the free energy. It is the equivalent of the Matone relation for SUSY gauge theories \cite{Matone1995}. The normalization condition expands in $\e$, providing refined approximations for the saddle point $q^\ast=q^\ast_0(\a)+\e q^\ast_1(\a)+\cdots$.

In order to investigate the subleading orders, we need to introduce some notation for the large $N$ expansion of the canonical model at $\e=\a/N$ with $\a$ fixed,
\begin{equation}\label{int}
\FC(N,\a/N)=\dfrac1{N}\log\ZC(N,\a/N)=\sum_{n=0}^\infty N^{-n}\FC^{(n)}(\a).
\end{equation}
Let us emphasize that this expansion is different from the standard topological expansion at fixed $\e$. It is the reason why the one-loop term in \ref{ZC_1} is not only given by the determinant but also contains corrective terms to the action. The inverse discrete Laplace transform \ref{Z_trans} with the constraint $\e N=\a$ specializes to
\begin{equation}\label{Ztrans}
e^{N\FC(N,\a/N)}=\dfrac{\a^NN!}{N^N}\oint_0{\dfrac{dq}{2i\pi q}e^{-N\left[\log q-\a^{-1}\FGC(q,\a/N)\right]}}.
\end{equation}
At sub-leading order, this integral is approximately equal to 
\begin{equation}\label{Ztrans_2nd_order}
e^{N\FC(N,\a/N)}=\a^Ne^{-N}\dfrac1{iq^\ast\sqrt{d}}e^{-N\left[\log q^\ast-\a^{-1}\FGC(q^\ast,\a/N)\right]}+O(1/N),
\end{equation}
with $q^\ast(\a)$ solution of the normalization condition \ref{qdq}, and
\begin{equation}
d=\left.\p_{q}^2\left[\log q-\a^{-1}\FGC(q)\right]\right|_{q=q^\ast}.
\end{equation}
This quantity $d$ can be expressed in terms of the norm $n$ of the two points grand canonical density $\brho(x,y)$, defined in \ref{def_GC_dens}, using a formula derived in appendix \refOld{App_qdq},
\begin{equation}\label{qdq2}
n=\int{\brho(x,y)dxdy}=(q\p_q)^2\FGC(q).
\end{equation}
At the saddle point, we have $n=-(q^\ast)^2\a d$. Expanding \ref{Ztrans_2nd_order} in $\e$, we obtain at second order the following relation between free energies,
\begin{equation}\label{rel_F1}
\FC^{(1)}=\FGC^{(1)}(q^\ast_0)-\hf\log(n_0/\a).
\end{equation}
To retrieve the equality previously observed among the free energies at subleading order, we have to assume that the norm $n_0$ of $\brho_0(x,y)$ is equal to $\a$ at the saddle point. It implies that the tree-level propagator $Y_0(x,y)$, which is the generating function of bi-rooted trees, has a vanishing norm (see \ref{brho_Y_2pts} below). It is however possible that we missed a factor in our treatment of the canonical partition function, in particular when we discarded the zero-mode in appendix \refOld{App_entropic}. This is why we will remain cautious and keep the critical value of $n_0$ arbitrary in the following.

\paragraph{Density and dressed vertex} The comparison of the effective actions led us to propose an identification between the tree-level dressed vertex of the grand canonical model and the large $N$ eigenvalue density associated to the canonical model. This identification can also be derived by general considerations involving the discrete Laplace transform. It will be done here in two steps. First we have to relate the dressed vertex $Y(x)$ to a grand canonical density $\brho(x)$. Then, we will exploit the inverse Laplace transformation to deduce an equality between canonical and grand canonical densities at first order.

To complete our program, we need to define the grand canonical vev of an operator $\Op(x)$,
\begin{equation}
\la\Op(x)\ra=\dfrac1{\ZGC(\bq)}\sum_{N=0}^\infty\dfrac{\bq^N}{N!}\ZC(N)\laN\Op(x)\raN.
\end{equation}
It is expressed in terms of the canonical vevs,
\begin{equation}
\laN\Op(x)\raN=\dfrac1{\ZC(N)}\int_{\mathbb{R}^N}{\Op(x)\prod_{i=1}^NQ(\phi_i)\dfrac{d\phi_i}{2i\pi}\prod_{\superp{i,j=1}{i<j}}^NK(\phi_i-\phi_j)},
\end{equation}
where the operator depends on $N$ fields in a permutation invariant manner. We focus on the density operator, and consider the sourced partition function
\begin{equation}
\ZGC[J]=\ZGC\la\exp\left(\int{dx J(x)\CD(x)}\right)\ra,\quad \CD(x)=\sum_i\d(x-\phi_i).
\end{equation}
The grand canonical density is defined as
\begin{equation}\label{def_brho_II}
\brho(x)=\e\la\CD(x)\ra=\e\left.\dfrac{\d\log\ZGC[J]}{\d J(x)}\right|_{J=0}.
\end{equation}
Introducing the source term $J$ in the partition function corresponds to replace the potential by $Q(\phi)\to e^{J(\phi)}Q(\phi)$, as can be seen from 
\begin{equation}
\ZGC[J]=\ZGC\la\prod_ie^{J(\phi_i)}\ra.
\end{equation}
Thus, the Mayer expansion also applies to the sourced quantity, leading to \ref{sum_clusters} with $e^{J(\phi_i)}$ inserted into the product over vertices. From this expression, we compute the derivative
\begin{equation}
\left.\dfrac{\d\log\ZGC[J]}{\d J(x)}\right|_{J=0}=\sum_{l=0}^\infty{\bq^l\sum_{C_l}\dfrac{l}{\s(C_l)}\dfrac{Q(x)}{2i\pi}\int{\prod_{i\in V(C_l)\setminus\{x\}}Q(\phi_i)\dfrac{d\phi_i}{2i\pi}\prod_{<ij>\in E(C_l)}\e f(\phi_{ij})}}.
\end{equation}
The identity \ref{sum_sym} demonstrated in appendix \refOld{App_qdq} allows to replace the clusters summation by a summation over rooted clusters. In doing so, we obtain exactly the dressed vertex
\begin{equation}
\left.\dfrac{\d\log\ZGC[J]}{\d J(x)}\right|_{J=0}=\dfrac1{2i\pi\e} Y(x),
\end{equation}
and we deduce from \ref{def_brho_II} the identity $Y(x)=2i\pi\brho(x)$ at the level of one marked point. It is easy to check that this identification is compatible with the property \ref{qdq} by computing the norm of the density $\brho(x)$ from the definition \ref{def_brho_II}.

It remains to take the Laplace transform. The canonical density is by definition
\begin{equation}
\rho(x)=\dfrac1{N}\laN\CD(x)\raN,
\end{equation}
it reduces to the collective field at large $N$, $\rho(x)\simeq\rho_0(x)$. Hence, $\ZGC\brho$ is related to $\a\ZC\rho$ by a discrete transformation similar to \ref{def_CZ}. Using a saddle point technique, we find at subleading order the relation $\brho_0(x)=\a\rho_0(x)$, in agreement with the proposed identification between $Y_0$ and $\rho_0$. It is also possible to derive this relation considering the inverse Laplace transform of the sourced partition function. It implies a Legendre relation of the type \ref{Legendre_FE} among sourced free energies. Taking the functional derivative with respect to the source $J$, we recover the relation between one-point density. One has to be careful because the saddle point depends on the source. But, contrary to the case of 2-points densities treated below, the dependence vanishes here.

\paragraph{Higher point densities and cluster generating functions} The previous argument generalizes to a higher number of marked vertices and multi-points densities. The two points grand canonical density defined as the connected correlator\footnote{Since the partition function behave at small $\e$ as $\ZGC\sim e^{\e^{-1}\FGC^{(0)}}$, we have the factorization property \cite{Drouffe1983},
\begin{equation}
\la\CD(x)\CD(y)\ra=\la\CD(x)\ra\la\CD(y)\ra+\la\CD(x)\CD(y)\ra_c\quad\text{with}\quad \dfrac{\la\CD(x)\CD(y)\ra_c}{\la\CD(x)\ra\la\CD(y)\ra}=O(\e).
\end{equation}
It ensures that $\brho(x,y)=O(1)$.}
\begin{equation}\label{def_GC_dens}
\brho(x,y)=\left.\dfrac{\d^2\FGC[J]}{\d J(x)\d J(y)}\right|_{J=0}=\e\la\CD(x)\CD(y)\ra_c,
\end{equation}
relates to the full propagator $Y(x,y)$, generating function of bi-rooted trees, as
\begin{equation}\label{brho_Y_2pts}
\brho(x,y)=\dfrac1{(2i\pi)^2}Y(x,y)+\dfrac1{2i\pi}\d(x-y) Y(x).
\end{equation}
This identity is obtained by taking the second derivative of the sourced free energy $\FGC[J]$. The additional term in the RHS of this relation corresponds to coinciding eigenvalues in the decomposition
\begin{equation}
\e^{-1}\brho(x,y)=\la\sum_{i\neq j}\d(x-\phi_i)\d(y-\phi_j)\ra_c+\la\sum_i\d(x-\phi_i)\d(y-\phi_i)\ra.
\end{equation}
The second term in the RHS produces a delta function of $x-y$ times the one-point density, and $Y(x,y)$ corresponds to the non-diagonal terms.

The second part of the argument exploits the fact that the sourced partition functions are also related through a discrete Laplace transform. But now the saddle point $q^\ast$ depends on the source $J$, for instance
\begin{equation}
\left.\dfrac{\d q^\ast}{\d J(x)}\right|_{J=0}=-\dfrac{q^\ast}{n}\int{\brho(x,y)dy}.
\end{equation}
At leading order, the sourced free energies satisfy the equation \ref{Legendre}. Taking twice the derivative with respect to the source, we obtain the relation between two points densities at first order,
\begin{equation}\label{rel_2pts}
\a\rho_0(x,y)=\brho_0(x,y)-\dfrac1{n_0}\int{\brho_0(x,u)du}\int{\brho_0(y,v)dv},
\end{equation}
where $\rho_0(x,y)$ is the leading order of the canonical two points connected density
\begin{equation}
\rho(x,y)=\left.\dfrac{\d^2\FC[J]}{\d J(x)\d J(y)}\right|_{J=0}=\dfrac1N\laN\CD(x)\CD(y)\raN_c.
\end{equation}
The second term in \ref{rel_2pts} is due to the dependence of the saddle point in the source. This expression is compatible with the requirement of vanishing norm for the connected correlator $\rho_0(x,y)$.

At higher points, we expect relations similar to \ref{brho_Y_2pts} to hold between multi-rooted clusters generating functions and grand canonical densities. They can be derived by performing higher derivations of the free energy with respect to the source. On the other hand, the relation between canonical and grand canonical densities becomes increasingly complicated and cannot be worked out easily using this method, even at the planar order.

\section{Loop equations}
In the previous section, we have compared canonical and grand canonical models at the level of free energies. We have shown how to recover the collective field theory description of the canonical model from the cluster expansion of the grand canonical partition function. This comparison was restricted to the two first orders in large $N$ and small $\e$. On the canonical side, it is possible compute higher order terms by employing the recursive technique of loop equations. This technique, originally developed for matrix models, has recently been extended to a large class of models to which $\ZC$ belongs \cite{Borot2013}.\footnote{Our model corresponds to the special case $\b=0$ and $\rho=1$ in their notations. For this value of $\b$, and depending on the explicit expression for the kernel, some of the assumptions considered in their paper may not be satisfied. This would have to be checked case by case.} Our goal in this section is to map these loop equations to similar relations among objects pertaining to the cluster expansion. These objects are the $n$-points $Y$-functions, generating functions of $n$-rooted clusters. We have already encountered the cases $n=1$ and $n=2$, corresponding respectively to the dressed vertex $Y(x)$ and the propagator $Y(x,y)$.

Loop equations for the canonical densities are obtained in the following manner. First, the invariance of the measure allows to write a set of linear relations among (non-connected) correlators. These correlators are decomposed into connected parts. The connected correlators involved are resolvents, i.e. multiple Cauchy transforms of the densities. As such, they have a branch cut along the support $\G$ of the densities in each of their variables. Taking the discontinuities of the previous equations, we are able to derive a set of coupled integral equations among densities. These equations can be expanded in large $N$, and solved recursively. The recursion involves both the genus, that is the order in $N^{-1}$ and the number of points. At this level, loop equations also depends on the derivative of densities. They can be integrated with a little bit of algebra. The resulting 'primitive' equations no longer contain the densities derivative. In the process, a constant of integration appears. It is fixed by imposing a vanishing norm to the $n$-points densities with $n>1$.

Grand-canonical densities also obey the canonical loop equations. Indeed, those equations are linear in the (non-connected) canonical correlators, and valid for any $N$. They can be summed over $N$ with appropriate coefficients to produce equations among grand canonical correlators. Next, these correlators are decomposed into connected parts. We must emphasize that the connected grand canonical correlators are no-longer the discrete Laplace transform of canonical ones. These connected correlators are also the resolvents associated to the multi-points grand canonical densities. For $\e$ infinitesimal, these densities are assumed to be continuous on a connected support, just like the canonical ones. The discontinuity process still works, leading to the same 'derivative' loop equations. Densities are then expanded in $\e$, which plays a role equivalent to the the large $N$ topological expansion for the canonical model. After integration, we recover the same integral equations, but with different constant of integrations since grand canonical and canonical densities have a different norm.

In this section, the strategy is as follows. We first provide the derivation of the canonical loop equation, and re-write them in the integrated form. Then, we compare this equation with a relation among $Y$-functions derived using the Mayer expansion. We deduce from the relation between $Y$ and $\brho$ that this density obey the integrated canonical loop equation. We conclude that the $Y$-function relations are the equivalent of loop equations. Finally, a technique to derive the loop equation for grand canonical densities is presented in subsection \refOld{sec_res_equ}.

\subsection{One-point density at leading order and rooted trees}
The simplest loop equation is derived from the identity
\begin{equation}
0=\sum_{k=1}^N\int{\prod_{i=1}^Nd\phi_i\dfrac\p{\p\phi_k}\left[\dfrac1{z-\phi_k}\prod_{i=1}^N \dfrac{Q(\phi_i)}{2i\pi}\prod_{\superp{i,j=1}{i<j}}^NK(\phi_{ij})\right]}.
\end{equation}
It produces an equation satisfied by the resolvent $W(z)$ which is the Cauchy transform of the density $\rho(x)$,
\begin{equation}
W(z)=\int{\dfrac{\rho(x)dx}{z-x}}=\dfrac1{N}\laN\sum_{i=1}^N\dfrac1{z-\phi_i}\raN.
\end{equation}
We will also need to introduce an auxiliary quantity $P(z)$ defined as
\begin{equation}
P(z)=\dfrac1{N}\laN\sum_{k=1}^N\dfrac{V'(z)-V'(\phi_k)}{z-\phi_k}\raN,\quad V(z)=\log \dfrac{Q(z)}{2i\pi},
\end{equation}
with $V(z)$ the standard 'matrix model' potential. Then, the first loop equation takes the form
\begin{equation}\label{1st_leq}
P(z)=-W'(z)+V'(z)W(z)+N\int_{\mathbb{R}^2}{\dfrac{k(x-y)}{z-y}\left(\rho(x)\rho(y)+\dfrac1N\rho(x,y)\right)dxdy}.
\end{equation}
with the shortcut notation $k(x)=\p_x\log K(x)$ for the logarithmic derivative of the kernel.

We have assumed that the eigenvalues condense on the support $\G$ of $\rho$ in the large $N$ limit. It implies that $W(z)$ has a branch cut on $\G$, with a discontinuity given by $-2i\pi\rho(x)$. On the other hand, by construction $P(z)$ is not singular over $\G$. Thus, taking the discontinuity of the loop equation \ref{1st_leq} over $\G$ allows to eliminate $P(z)$ and write an equation involving only densities,
\begin{equation}\label{leq1}
\rho'(x)=\p_x\log(Q(x))\rho(x)+\int_\mathbb{R}{k(x-y)\left(N\rho(x)\rho(y)+\rho(x,y)\right)dy}.
\end{equation}
The next step is to expand the density at large $N$, we denote $\rho_n$ the term of order $O(N^{-n})$. At the first order, the dependence in the two points density drops, and we find an integral equation for $\rho_0$ which is precisely the equation of motion \ref{eom_SC} derived from the canonical action $\SC$. Integrating once, we recover the integral equation \ref{int_equ} obeyed by $Y_0$,
\begin{equation}\label{leq1_0}
\log\left(\dfrac{2i\a\pi\rho_0(x)}{Q(x)}\right)=\a\int{f(x-y)\rho_0(y)dy}+\g_0,
\end{equation}
provided we choose the integration constant $\g_0$ to be $\log q$. A priori, the unit norm constraint over the density should fix this integration constant. However, it is very non-trivial to impose this condition in practice due to the complicated form of the integral equation. At the saddle point, $Y_0(x)=2i\pi\brho_0(x)=2i\pi\a\rho_0(x)$ and $\g_0=\log q_0^\ast(\a)$.

\subsection{Two-points density at leading order and bi-rooted trees}
The subleading order of the first loop equation contains the two points density at first order $\rho_0(x,y)$. To compute this quantity, we need a second loop equation, derived from the identity
\begin{equation}
0=\sum_{k=1}^N\int{\prod_{i=1}^Nd\phi_i\dfrac\p{\p\phi_k}\left[\dfrac1{z-\phi_k}\sum_{l=1}^N\dfrac1{w-\phi_l}\prod_{i=1}^N\dfrac{Q(\phi_i)}{2i\pi}\prod_{\superp{i,j=1}{i<j}}^NK(\phi_{ij})\right]}.
\end{equation}
It provides an equation satisfied by the two points resolvent $W(z,w)$ and involving an auxiliary quantity $P(z,w)$,
\begin{equation}
W(z,w)=\dfrac1{N}\laN\sum_{i,j=1}^N\dfrac1{z-\phi_i}\dfrac1{w-\phi_j}\raN_c,\quad P(z,w)=\dfrac1N\laN\sum_{i=1}^N\dfrac{V'(z)-V'(\phi_i)}{z-\phi_i}\sum_{j=1}^N\dfrac1{w-\phi_j}\raN_c.
\end{equation}
This equation also involves the three-points density, but this dependence drops at leading order. The first loop equation \ref{1st_leq} can be used to simplify the result, which gives
\begin{align}
\begin{split}
P(z,w)=&-\p_zW(z,w)+\p_w\left(\dfrac{W(z)-W(w)}{z-w}\right)+V'(z)W(z,w)\\
&+N\int{\dfrac{k(x_1-x_2)}{(z-x_2)(w-x_3)}\left[\rho(x_1,x_3)\rho(x_2)+\rho(x_2,x_3)\rho(x_1)+\dfrac1N\rho(x_1,x_2,x_3)\right]dx_1dx_2dx_3}.
\end{split}
\end{align}
Just like $P(z)$, $P(z,w)$ has not branch on $\G$ for its variable $z$, and will be eliminated by taking the discontinuity of the equation. In this process, the difference of resolvents must be regularized at coincident values as follows,\footnote{To derive this contact term, we take a test function $r(x)$ regular on the branch cut $\G$, and consider
\begin{equation}
\int_\G{r(y)dy \Disc_y\Disc_x \dfrac{W(x)-W(y)}{x-y}}=2i\pi\rho(x)\oint_\G{\dfrac{r(y)}{x-y}dy}=-(2i\pi)^2\rho(x)r(x),
\end{equation}
since $x$ belongs to the integration contour.}
\begin{equation}
\dfrac{W(x)-W(y)}{x-y}\to-(2i\pi)^2\d(x-y)\rho(x).
\end{equation}
Taking the discontinuity over the variable $z$ and $w$, extracting the first order and integrating once, we get
\begin{equation}\label{equ_r0_xy}
\rho_0(x,y)=\rho_0(x)\d(x-y)+\a\rho_0(x)\int{duf(x-u)\rho_0(u,y)}+\g_{00}(y)\rho_0(x),
\end{equation}
with the integration constant $\g_{00}(y)$ that may depend on $y$. This degree of freedom is fixed by imposing that $\rho_0(x,y)$ is a symmetric function of its parameters, and has zero norm since it is a connected density.

We would like to recover the loop equation \ref{equ_r0_xy} using the Mayer expansion. According to our previous discussion in subsection \refOld{sec_GC_VS_C}, this equation should be obeyed by the tree-level propagator $Y_0(x,y)$, generating function of bi-rooted tree. Let us recall its definition,
\begin{equation}
Y_0(x,y)=Y_0(x)\sum_{l=1}^\infty\bY_l(x,y),\quad \bY_{l+1}(x,y)=Y_0(y)\int{\prod_{i=1}^lY_0(\phi_i)\dfrac{d\phi_i}{2i\pi} f(x-\phi_1)f(\phi_{12})\cdots f(\phi_{l-1\ l}) f(\phi_l-y)},
\end{equation}
where $Y_0(x)\bY_{l+1}(x,y)$ is the generating function of bi-rooted trees such that the roots are connected through a chain of $l$ intermediate vertices (and $l+1$ links). We should also supply the definition of the first member of this set of functions, $\bY_1(x,y)=f(x-y)Y_0(y)$, obtained when the roots are directly connected.\footnote{Note that the free energy at subleading order can be expressed using $\bY_l(x,y)$ if we merge the two roots in order to build a cycle,
\begin{equation}
\FGC^{(1)}(\bq)=\sum_{l=3}^\infty{\dfrac1{2l}\int{\dfrac{dx}{2i\pi}\bY_l(x,x)}}.
\end{equation}} Contrary to the functions $\bY_l(x,y)$, $Y_0(x,y)$ is a symmetric function of $x$ and $y$. The functions $\bY_l$ obey an obvious recursion relation that is interpreted as attaching to the vertex $x$ a new rooted vertex $z$,
\begin{equation}
\bY_{l+1}(z,y)=\int{f(z-x)\bY_l(x,y)Y_0(x)\dfrac{dx}{2i\pi}}.
\end{equation}
In this process, $x$ is still 'marked' in the sense that it is determined uniquely being the first vertex attached to $z$ on the path to $y$, but we will not consider it as a 'root' anymore, preferring the endpoint $z$. Summing over $l$, we deduce the integral equation obeyed by $Y_0(x,y)$,
\begin{equation}\label{equ_Y0_xy}
Y_0(x,y)=f(x-y)Y_0(x)Y_0(y)+Y_0(x)\int{f(x-z)Y_0(z,y)\dfrac{dz}{2i\pi}}.
\end{equation}
Making use of the relation \ref{brho_Y_2pts} between the grand canonical two points density $\brho(x,y)$ and the propagator $Y(x,y)$, we deduce that at first order in $\e$, $\a^{-1}\brho_0(x,y)$ obey the integrated loop equation \ref{equ_r0_xy} with a vanishing integration constant $\g_{00}(y)=0$,
\begin{equation}\label{equ_br0_xy}
\brho_0(x,y)=\brho_0(x)\d(x-y)+\brho_0(x)\int{duf(x-u)\brho_0(u,y)}.
\end{equation}

The relation \ref{rel_2pts} between $\rho_0(x,y)$ and $\brho_0(x,y)$ is compatible with the loop equations \ref{equ_r0_xy} and \ref{equ_br0_xy}. This can be shown by taking the $q$-derivative of the equation \ref{int_equ} satisfied by $\brho_0(x)=Y_0(x)/2i\pi$,
\begin{equation}
q\p_q\brho_0(x)=\brho_0(x)+\brho_0(x)\int{f(x-u)q\p_q\brho_0(u)du},\quad\int{\brho_0(x,y)dy}=q\p_q\brho_0(x).
\end{equation}
The second equality is a consequence of \ref{B7} and \ref{brho_Y_2pts}. We deduce the expression of $\g_{00}(y)$ at the saddle point,
\begin{equation}
\g_{00}(y)=-\dfrac1{n_0}\int{\brho_0(y,u)du}.
\end{equation}

\subsection{One-point density at subleading order and rooted 1-cycles}
To obtain the equation satisfied by the genus one correction to the 1-point density $\rho_1(x)$, we examine the first loop equation \ref{leq1} at subleading order. Again, the result can be simplified using the first order result \ref{leq1_0}, and integrated, leading to 
\begin{equation}\label{leq1_II}
\dfrac{\rho_1(x)}{\rho_0(x)}=\a\int{f(x-y)\rho_1(y)dy}-\hf\a^2\int{f(x-y)^2\rho_0(y)dy}+s(x),
\end{equation}
where $s(x)$ contains the contribution of the two points density,
\begin{equation}\label{s_d}
s'(x)=\dfrac\a{\rho_0(x)}\int{f'(x-y)\rho_0(x,y)dy}.
\end{equation}
The equation \ref{equ_r0_xy} obtained upon the 2-points density can be used to simplify the expression \ref{s_d} and integrate it, leading to the following loop equation for $\rho_1(x)$,\footnote{The equation \ref{equ_r0_xy} has originally be obtained in the form
\begin{equation}\label{equ_r0_xy_d}
\p_x\left(\dfrac{\rho_0(x,y)}{\rho_0(x)}\right)=\d'(x-y)+\a\int{duf'(x-u)\rho_0(u,y)}.
\end{equation}
Integrating this expression multiplied by $f(x-y)$ over $y$, and then using the primitive relation \ref{equ_r0_xy} to simplify the result, we obtain the identity
\begin{equation}
\int{f(x-y)\p_x\left(\dfrac{\rho_0(x,y)}{\rho_0(x)}\right)dy}=\int{f'(x-y)\left[\dfrac{\rho_0(x,y)}{\rho_0(x)}-\g_{00}(y)\right]dy},
\end{equation}
since $f'(0)=0$. This identity is then plugged into the expression \ref{s_d} of $s'(x)$.} 
\begin{equation}\label{leq1_III}
\dfrac{\rho_1(x)}{\rho_0(x)}=\a\int{f(x-y)\rho_1(y)dy}-\hf\a^2\int{f(x-y)^2\rho_0(y)dy}+\hf\dfrac\a{\rho_0(x)}\int{f(x-y)\left[\rho_0(x,y)+\g_{00}(y)\rho_0(x)\right]dy}+\g_1,
\end{equation}
where the integration constant $\g_1$ is fixed by a normalization condition.

\begin{figure}[!t]
\centering
\includegraphics[width=10cm]{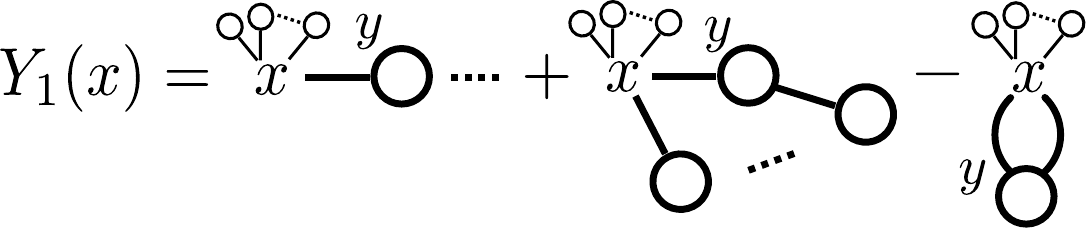}
\caption{Equation satisfied by the one-cycle dressed vertex, empty circles represent tree-level dressed vertices.}
\label{Y1_leq}
\end{figure}

The integrated loop equation \ref{leq1_III} we have obtained for $\rho_1(x)$ should be compared to the equation satisfied by $Y_1(x)$, the generating function of rooted clusters with exactly one cycle,
\begin{equation}\label{equ_Y1}
Y_1(x)=Y_0(x)\int{f(x-y)Y_1(y)\dfrac{dy}{2i\pi}}+\hf\int{Y_0(x,y)f(x-y)\dfrac{dy}{2i\pi}}-\hf Y_0(x)\int{f(x-y)^2Y_0(y)\dfrac{dy}{2i\pi}}.
\end{equation}
In this expression, represented graphically on figure \refOld{Y1_leq}, the first term corresponds to the case where $x$ does not belong to the cycle. Hence, there is a vertex $y$, directly linked to $x$ such that if we remove this link, the cycle is present in the cluster rooted by $y$. In the second term, $x$ directly belongs to the cycle. In this case, we choose a vertex $y$ from the cycle and directly connected to $x$. Cutting the link $x-y$, we obtain a bi-rooted tree. In the process, we gain a symmetry factor $1/2$ due to the choice of $y$. Finally, the third term correspond to trees of $Y_0(x,y)$ for which $x$ and $y$ are directly related. For those clusters, $x$ and $y$ cannot get an extra link, and their contribution must be withdrawn from the previous term. Comparing \ref{leq1_III} and \ref{equ_Y1}, we deduce that $\brho_1(x)=Y_1(x)/2i\pi$ satisfies the loop equation \ref{leq1_III} with vanishing integration constants $\g_1=\g_{00}(y)=0$.

\subsection{Yet another way to derive grand canonical loop equations}\label{sec_res_equ}
Another possibility to establish grand canonical loop equations is to start from the definition of the density $\brho(x)$ and make use of the $\d$-function to fix one of the integration variables. The canonical correlator of $N$ variables reduces to a correlator of $N-1$ variables, and after summation over $N$ we obtain an equation satisfied by the grand canonical density,
\begin{equation}\label{GC_leq1}
2i\pi\brho(x)=qQ(x)\la\prod_iK(x-\phi_i)\ra.
\end{equation}
At first order in $\e$, it is possible to use the factorization property
\begin{equation}
\la\prod_{i=1}^kK(x-\phi_i)\ra\simeq \la e^{\e\sum_i f(x-\phi_i)}\ra\simeq \exp\left(\e\la\sum_if(x-\phi_i)\ra\right).
\end{equation}
The sum over $\phi_i$ can be replaced by an integral over density $\rho(x)$. Expanding \ref{GC_leq1} in $\e$, and keeping only the first order, we deduce that $\brho_0(x)$ satisfies the integral equation \ref{int_equ} with $Y_0(x)=2i\pi\brho_0(x)$.

At the subleading order in $\e$, the factorization property becomes
\begin{equation}
\la\prod_iK(x-\phi_i)\ra=e^{\e\la\sum_if(x-\phi_i)\ra}\left[1-\hf\e^2\la\sum_if(x-\phi_i)^2\ra+\hf\e^2\la\sum_{i,j}f(x-\phi_i)f(x-\phi_j)\ra_c+O(\e^2)\right].
\end{equation}
In the RHS bracket, the first term comes from the expansion of $\log K$, the last term is the first order correction to the factorization property. Replacing sum over variables $\phi_i$ with densities, and expanding in $\e$, we get at the second order an equation satisfied by the subleading correction to $\brho(x)$,
\begin{equation}\label{equ_brho1}
\dfrac{\brho_1(x)}{\brho_0(x)}=\int{f(x-u)\brho_1(u)du}-\hf\int{f(x-u)^2\brho_0(u)du}+\hf\int{f(x-u)f(x-v)\brho_0(u,v)dudv},
\end{equation}
where we have used the first order equation to simplify the result. Using the equation \ref{equ_br0_xy} to deal with the last term, this loop equation reproduces the equation \ref{equ_Y1} obtained from the cluster expansion upon the identification $Y(x)=2i\pi\brho(x)$ and \ref{brho_Y_2pts} of the densities.

The same argument can be repeated for the two points density. Fixing an integration variable using the operator $\CD(x)$ in the two-point correlator, we find
\begin{equation}\label{GC_leq2}
\e\la\CD(x)\CD(y)\ra=\d(x-y)\brho(x)+\dfrac{qQ(x)}{2i\pi}\la\CD(y)\prod_iK(x-\phi_i)\ra,
\end{equation}
where the result has been simplified using the first loop equation \ref{GC_leq1}. From the factorization property
\begin{equation}
\la\CD(y)\prod_iK(x-\phi_i)\ra\simeq\la\prod_iK(x-\phi_i)\ra\left(\la\CD(y)\ra+\e\la\sum_if(x-\phi_i)\CD(y)\ra_c\right),
\end{equation}
we recover at subleading order in $\e$ the equation \ref{equ_br0_xy} satisfied by the two points density $\brho_0(x,y)$.

\section{Concluding remarks}
In these notes we compared the cluster expansion of a grand canonical model with the standard matrix model treatment of its canonical partition function. At tree level, the grand canonical free energy is given by the minimum of an effective action which is identical to the one provided by the collective field theory approach applied to the canonical model. The correspondence extends to the level of one-loop corrections, where the sum over one-cycle clusters reproduces the expansion of the Fredholm determinant computed from the integration over Gaussian fluctuations in the collective field theory. The matching of free energies can be explained by the discrete Laplace transform relating canonical and grand canonical models. Introducing a source term, we were also able to find the relation satisfied by the one-point and two points densities.

We continued with the study of canonical loop equations, and realized that a similar set of equations can be derived from the cluster expansions. Instead of $n$-point connected densities, these equations involve the generating functions of $n$-rooted clusters, denoted $Y$. Using these equations, and the relation between $Y$-functions and densities derived earlier, we verified that grand canonical densities also obey the canonical loop equations. It implies that the whole loop equation structure is present in the cluster expansion, and takes the form of graphical relations among clusters. Finally, we proposed a method to derive directly this set of loop equations within the grand canonical model.

Our study is restricted to the first two orders in the large $N$ and small $\e$ expansion. The generalization to higher orders still needs to be done, and the general form of loop equations to be worked out. Once the full set of equations identified, it may be possible to apply the topological recursion to the grand canonical model.

Another important point that remains is the description of instantons clustering relevant to the instanton partition function of $\mathcal{N}=2$ SUSY gauge theories. The dual description grand-canonical/canonical may allow a better understanding of this phenomenon. A possible application for this work could be the derivation of the subleading correction in $\e_2$ to the partition function, and the investigation of its presumed integrable properties. In this scope, it is tempting to assume that the instanton clustering in SYM is entirely described by the effective action \ref{action_NS}, and conjecture that the subleading order is given by the associated determinant,
\begin{align}
\begin{split}
e^{\CF_\text{NS}^{(1)}}&=\dfrac1{\sqrt{\det\left[\d(x-y)+(1-e^{\rho(x)})G(x-y)\right]}}e^{\frac12\int{dx\log\left(e^{\rho(x)}-1\right)}-\frac12\int{dx\log\rho(x)}}\\
&\times\exp\left(\hf G(0)\int{dx(1-e^{\rho(x)})}-\dfrac14\int{dxdy(1-e^{\rho(x)})(1-e^{\rho(y)})G(x-y)^2}\right).
\end{split}
\end{align}
This proposal is very naive, but it could be tested using the AGT correspondence with the \bens\ representation of Liouville correlators.

The grand canonical model we studied has a very specific form of interaction and may not be relevant to statistical systems. It would be interesting to consider more physical models. One may also wonder if the topological reduction employed in this context have a matrix model analogue. Nevertheless, the results presented here are very general and could be relevant for a large spectrum of problems. They have deep connections with integrable models and the TBA equation \cite{Basso2013}. They play a role in the computation of light-like Wilson loops at strong coupling in $\mathcal{N}=4$ SYM \cite{Basso2013a}. They may also be applied to the study of 3-points function of scalar operators in this theory \cite{Kostov2012,Foda2013}.

\section*{Acknowledgements}
I would like to thank Dima Volin, Yutaka Matsuo and Ivan Kostov for valuable discussions, and in particular Benjamin Basso for sharing his unpublished results. It is also a pleasure to acknowledge the hospitality of Ewha University at the occasion of the workshop ``Solving AdS/CFT '', and of CQUeST (Sogang U.) and Tokyo University where parts of this work has been done. I acknowledge the Korea Ministry of Education, Science and Technology (MEST) for the support of the Young Scientist Training Program at the Asia Pacific Center for Theoretical Physics (APCTP).

\appendix

\section{Demonstration of the tree level free energy formula}\label{App_Basso}
In this appendix, we give a demonstration for the formula \ref{Basso} where the sum over tree clusters $T_l$ is given by the difference of the two terms,
\begin{equation}\label{def_AB}
A=\int_\mathbb{R}{Y_0(x)\dfrac{dx}{2i\pi},\quad\text{and}\quad B=\hf\int_{\mathbb{R}^2}}{\dfrac{dx}{2i\pi}\dfrac{dy}{2i\pi}Y_0(x)Y_0(y)f(x-y)}.
\end{equation}
It is easy to see that both terms expand as a sum of tree clusters $T_l$, weighted as in \ref{sum_trees}, but with different symmetry factors. The second term $B$ contains trees with at least one link, associated to the function $f$ present in the integral \ref{def_AB}, which we call the main link. Such clusters are formed by gluing two trees along this main link. To all the terms in the expansion of $A$ and $B$ correspond a term of the summation \ref{sum_trees}. On the other hand, a single term in \ref{sum_trees} corresponds to many terms of $A$ and $B$ series, since a tree $T_l$ can be rooted from any of its vertices, leading to $l$ terms in $A$, and has $l-1$ links that can be associated to the main link of a $B$-term.\footnote{This is true up to identification of automorphic clusters, which is necessary to avoid over-counting. However, it turns out that this does not play any role in our cancellation argument.} The fact that the formula \ref{Basso} holds has to do with the property that a tree $T_l$ has exactly $l$ links for $l-1$ vertices.

The strategy we follow is to consider the terms in $A$ and $B$ cluster expansions that can be identified with a given cluster $T_l$ of the summation \ref{sum_trees}. The corresponding terms in $A$ are obtained by rooting the vertices of the cluster $T_l$. In the same way, terms from the $B$-expansion are derived from edging the links of the tree $T_l$. We then show that cancellation occurs between terms of $A$ and $B$ expansions due to the coincidence of symmetry factors. The remaining term provides the contribution of $T_l$ to the free energy with the correct symmetry factor.

\subsection{Example: chain of vertices}
It is better to understand what is going on over a few examples. Here we focus on linear trees, i.e. chains of vertices, denoted $R_l$. To facilitate the argument, vertices will be numbered according to their order in the chain, from left to right. We further call $i$th link the edge linking the vertices $i$ and $i+1$. These chains appear in the free energy expression \ref{sum_trees} with a factor $\s(R_l)=2$ corresponding to a reflexion symmetry.

\begin{figure}[!t]
\centering
\includegraphics[width=12cm]{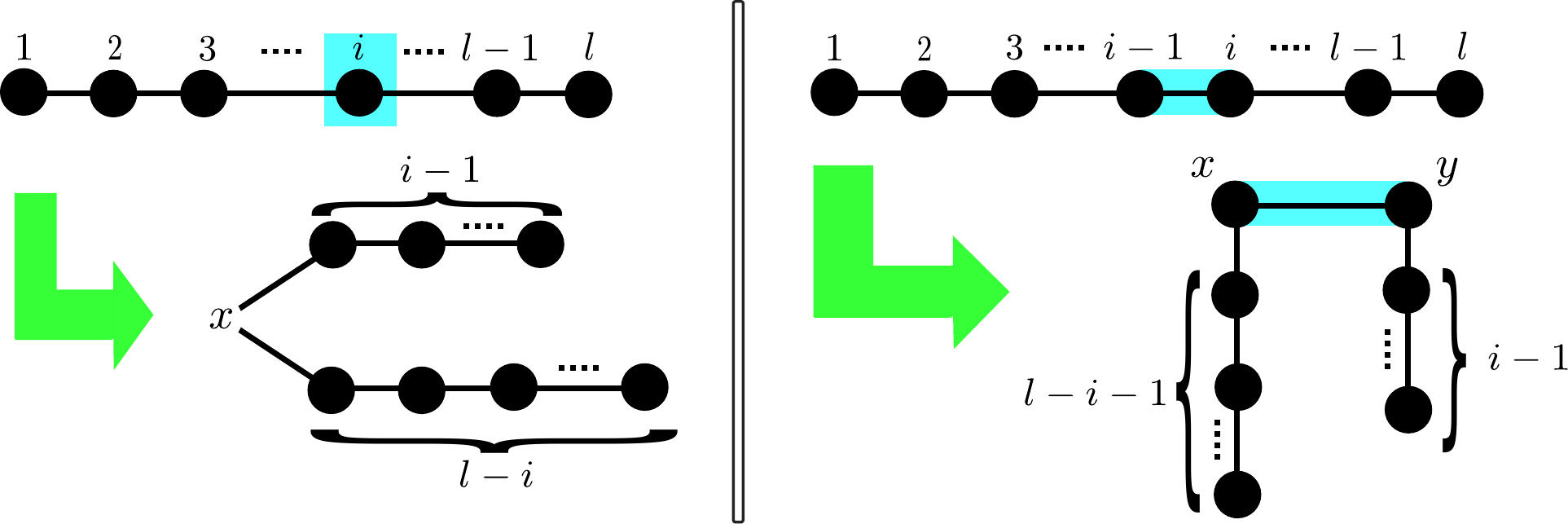}
\caption{Rooting a vertex (left) and edging a link (right) of a chain of $l$ vertices.}
\label{root_edge}
\end{figure}

We start with the case of a chain with odd length, $l=2k+1$. In the $A$-expansion, $R_l$ is associated to $k+1$ terms, corresponding to rooting the vertex $i$ (or $2k+1-i$) for $i=1\cdots k$ and the vertex $k+1$. This procedure is depicted in figure \refOld{root_edge} (left). The index $i$ runs from one to $k$ since rooting the vertex $i$ or $2k+1-i$ leads to the same tree. The first $k$th rooted trees obtained in this way have a symmetry factor of one since they are composed of two branches of length $i-1$ and $2k+1-i$ joining at the root. The remaining rooted tree, associated to the central vertex, has the symmetry factor $2$ since now both branches have the same length and can thus be exchanged.

Now, let us turn to the $B$-expansion. We consider the $B$ term having the $i$th link of $R_l$ as its main link. This $B$-term corresponds to a rooted chain of $i$ vertices glued to another rooted chain of $2k+1-i$ vertices through the main link. It is represented on figure \refOld{root_edge} (right). Its symmetry factor is one since the two trees on both sides of the main link have symmetry factor one and different length. We also have to take into account a 'chirality' factor of two counting the possibility of exchanging the two trees. This factor two is canceled by the factor $1/2$ in the definition of $B$. Here again we choose $i$ running only from one to $k$ since we get the same term after exchanging $i\to 2k+1-i$. Since the first $k$th $A$-terms are canceled by the $B$-terms, it remains only the contribution from the central vertex. As already mentioned, this contribution is weighted by $1/2$, thus providing the correct symmetry factor for the free energy cluster.


There is a lesson to learn from this example. As we will see later, it is possible to associate uniquely a link to each vertex but one by a recursive procedure. The corresponding terms in the $A$- and $B$-expansions cancel, and only the last vertex contributes. The rooted tree of this vertex has the same symmetry factor than the original cluster.

However a subtlety may appear. It is illustrated by our second example, the case of a chain with even length $R_{l=2k}$. Rooting the tree $R_l$ from the vertex $i$ (or $2k-i$) with $i=1\cdots k$, we obtain a tree with two branches of length $i-1$ and $2n-i$. These rooted trees have symmetry factor equal to one. Then, we consider the $i$-th edges with $i=1\cdots k-1$. They give $B$-terms consisting of two rooted chain, of length $i$ and $2k-i$, glued through the main link. The symmetry factor is one, and there is a chirality factor of two, again eliminated by the factor $1/2$. On the other hand, the central link, numbered $k$, is associated to two rooted trees of the same length $k$, and in such a case there is no chirality since right and left side of the $B$-terms are the same. We note that the $(k-1)$th first $A$-terms are eliminated by the chiral $B$-terms, and the last $A$-term gets subtracted by a half of its value, which corresponds to the $B$-term having the central edge as its main link. The difference $A-B$ thus reproduces the symmetry factor $2$ needed for the free energy.

\subsection{General case}
We now consider an arbitrary tree $T_l$ and associate recursively to each vertex a unique link, to which it connects directly, using the following procedure. First, the leaves, i.e. the vertices connected to only one other vertex, are naturally associated to the only link that end on them. Then, we remove those leaves in order to obtain a strictly smaller tree to which we repeat the procedure. At the end of the recursion, only two configurations may arise. In the first case, as for the odd chain, only a single vertex remains, all others are uniquely associated to a link. In the second case, typically for the even chain, a set of two vertices connected by a link remains. In this configuration, there is an ambiguity in the choice of the vertex to associate to the remaining link.

\begin{figure}[!t]
\centering
\includegraphics[width=6cm]{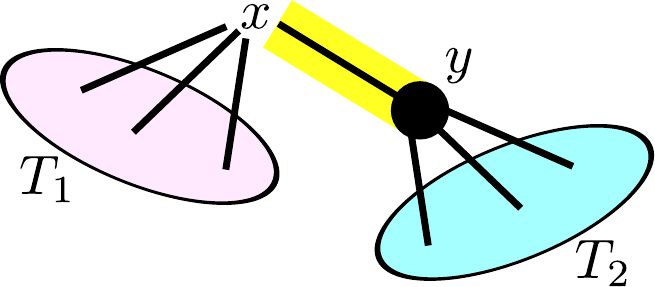}
\caption{Rooted tree obtained after rooting the vertex $x$, and its associated link (highlighted). It consists of two rooted subtrees, $T_1$ by $x$ and $T_2$ by $y$, linked by the $B$-term main edge.}
\label{general}
\end{figure}

As a second step, we argue that the $A$-term of a rooted vertex and the $B$-term of the cluster edged from the associated link cancel. To do so, we have to show that the rooted tree have the same symmetry factor than the tree with associated edge as main link. Let us take the $B$-term associated to an edge of the cluster $T_l$ which is not the final stage of the previous procedure. It consists of two rooted tree $T^{(1)}$ and $T^{(2)}$, with symmetry factors $\s_1$ and $\s_2$, such that the total factor is $\s_1\s_2$. We will see that these two composing trees are always different, so there is no symmetry enhancement. This object is chiral, which cancel the factor $1/2$ in front of the $B$ integral. It is associated to a rooted tree in the $A$-series which is displayed in figure \refOld{general}. By construction, the rooted vertex $x$ is an ending point of the main edge (in yellow on the figure). It is attached to the tree with the smallest deepness, i.e the smallest maximal distance between the leaves and the root.\footnote{The couples vertex-link can be labeled by an integer $n$ corresponding to the step of the recurrence at which vertex and link have been associated. Then, in the tree rooted at $x$, the root is attached to the main edge, and to a tree of deepness $n-1$. On the other side of the main edge lies a tree with strictly larger deepness.} The trees $T_1$ and $T_2$ cannot have the same deepness, otherwise the procedure of associating the vertex to an edge would not be unique. The main edge still increases the deepness of the deep tree on the right by one, and no symmetry enhancement can occur. Thus, the symmetry factor for the $A$-term is also $\s_1\s_2$. This shows the cancellation between $A$ and $B$ terms.

It remains to study the final stage of the procedure. The simplest case is when only one vertex remains. Then the symmetry factor of the rooted tree is equal to the one of the cluster $T_l$ we started from. Otherwise there would exist an automorphism exchanging the final vertex of $T_l$ with another one. But this is not possible as the root has been determined uniquely through the procedure described above.

Finally, we consider the case where one edge and two vertices remain at the end of the recursion. This edge connects two trees of a $B$-term with the same deepness, leaving to possibility of the two tree to be identical. If they are not identical, we can choose any of the two vertices to be the one associated with the remaining edge. Then, we can repeat the previous argument to show that the corresponding $A$ and $B$ terms cancel. No symmetry enhancement happens due to the fact that a node is added to one of the trees, thus incrementing its deepness. The rooted tree associated to the last vertex has again the same symmetry factor than the initial cluster. Otherwise it could be exchanged with the vertex we removed previously, meaning that the two trees where actually the same. The case where the two trees are identical has already been encountered in the example of the even chain. The $B$-term have symmetry factor $2\s_1^2$ where $\s_1=\s_2$ is the symmetry factor of the composing trees, and we included the $1/2$ prefactor which is no longer canceled by the chirality. The two vertices related to this edge leads to the same rooted tree, which has the symmetry factor $\s_1^2$. Taking the difference $A-B$, we get $1/\s_1^2-1/2\s_1^2=1/2\s_1^2$ which is exactly the symmetry factor of the initial cluster $T_l$, the factor of two taking into account the possibility of reflexion with respect to the final edge.

\section{Derivatives of the free energy from Mayer expansion}\label{App_qdq}
In this appendix, we give a proof of the formulas \ref{qdq} and \ref{qdq2}. First, examine the action of $q\p_q$ on the grand canonical free energy expressed as a sum over clusters as in \ref{sum_clusters},
\begin{equation}
q\p_q\FGC=\e\sum_{l=1}^\infty{l\bq^l\sum_{C_l}\dfrac1{\s(C_l)}\int{\prod_{i\in V(C_l)}Q(\phi_i)\dfrac{d\phi_i}{2i\pi}\prod_{<ij>\in E(C_l)}\e f(\phi_{ij})}}.
\end{equation}
The only effect of this operation is to multiply the cluster integrals by the number of vertices $l$. This expression should be compared to the cluster expansion of the integral of $Y(x)$. The integral of rooted clusters reproduces the cluster contributions of the free energy expansion, and we should only be concerned about the symmetry factor. To a given cluster $C_l$ of the free energy expansion corresponds $l$ rooted clusters $C_l^x$ obtained by rooting the vertices $x\in V(C_l)$. However some of these rooted clusters are identical. To avoid over-counting these terms, we separate the set of vertices into the sets $V_k(C_l)$ of vertices producing equivalent rooted clusters,
\begin{equation}
V(C_l)=\bigsqcup_{k}V_k(C_l).
\end{equation}
As already mentioned, the integral contributions of integrated rooted clusters producing the same cluster $C_l$ are equal and can be factorized. Thus, in order to prove \ref{qdq} we just need to establish
\begin{equation}\label{sum_sym}
\dfrac{l}{\s(C_l)}=\sum_k\dfrac1{\s(C_l^{x_k})},\quad x_k\in V_k(C_l).
\end{equation}

Now, let us discuss the group of automorphisms of the cluster $C_l$, denoted $\Aut(C_l)$. Consider one vertex $x\in C_l$ and the group of automorphism for the rooted cluster $C_l^x$. It is obvious that this group $\Aut(C_l^x)$ is a subgroup of $\Aut(C_l)$ consisting of the automorphism of $C_l$ that leave the vertex $x$ invariant. Those groups are subgroups of the group of permutations for the set vertices $\S(V(C_l))\simeq\S_l$ and can be decomposed into a product of transpositions.

We also need a formal definition of $V_k(C_l)$. Two vertices $x$ and $y$ produce an identical rooted cluster if and only if there exists an automorphism of $C_l$ mapping one into the other. Suppose we take an element $x_k\in V_k(C_l)$, then this set is the orbit of $x_k$ under the group of automorphisms,
\begin{equation}
V_k(C_l)=\{y\in V(C_l) / \exists g\in\Aut(C_l) / g.y=x_k\}=\{y\in V(C_l) / \exists g\in\Aut(C_l) / y=g.x_k\}.
\end{equation}
Note also that the groups of automorphisms for two vertices from $V_k(C_l)$ are isomorphic, $\Aut(C_l^x)\simeq\Aut(C_l^y)$, although they have a different representation on the cluster $C_l$.

Let $g\in\Aut(C_l)$ and $x\in V(C_l)$. There is a unique $k$ such that $x\in V_k(C_l)$. From the definition of $V_k(C_l)$, $g.x$ also belongs to $V_k(C_l)$ and we denote this element $x_k$. By construction $\t_{xx_k}g$ leaves the vertex $x$ invariant since $\t_{xx_k}$ is the transposition that exchanges $x$ and $x_k$. Therefore it is an element of $\Aut(C_l^x)$ that we denote $h$, and we have $g=\t_{xx_k} h$. It means that given a vertex $x$, any automorphism $g$ can be decomposed uniquely into its action on $x$, given by $\t_{xx_k}$ and another automorphism that leaves $x$ invariant, namely $h$.\footnote{Unicity. Let us suppose that there exists $\tilde{y}_k\in V_k(C_l)$ with $\tilde{y}_k\neq y_k$ and $\tilde{h}\in\Aut(C_l^x)$ such that $g=\t_{xy_k}h=\t_{x\tilde{y}_k}\tilde{h}$. It implies that $\t_{x\tilde{y}_k}\t_{xy_k}=h\tilde{h}^{-1}\in\Aut(C_l^x)$ in contradiction with the fact that $\t_{x\tilde{y}_k}\t_{xy_k}=(x\ y_k\ \tilde{y}_k)\notin\Aut(C_l^x)$.} We deduce
\begin{equation}
|\Aut(C_l)|=|V_k(C_l)|\times |\Aut(C_l^{x_k})|\quad \forall k, x_k\in V_k(C_l),
\end{equation}
which implies \ref{sum_sym}.

The demonstration remains valid if we replace the cluster $C_l$ by a rooted cluster $C_l^x$ and the rooted clusters $C_l^{x_k}$ with bi-rooted ones $C_l^{x,x_k}$, and \ref{sum_sym} becomes
\begin{equation}
\dfrac{l-1}{\s(C_l^x)}=\sum_k\dfrac1{\s(C_l^{x,x_k})},
\end{equation}
since we now have only $l-1$ unmarked vertices. We deduce the following relation between rooted and bi-rooted generating functions,
\begin{equation}\label{B7}
q\p_qY(x)=\int{Y(x,y)\dfrac{dy}{2i\pi}}+Y(x),
\end{equation}
which implies \ref{qdq2}. Similar formulas can be obtained for a higher number of roots, 
\begin{equation}
q\p_qY(x_1,\cdots,x_n)=\int{Y(x_1,\cdots,x_n,y)\dfrac{dy}{2i\pi}}+nY(x_1,\cdots,x_n).
\end{equation}
By recursion, it implies
\begin{equation}
q^n\p_q^n\FGC=\int{Y(x_1,\cdots,x_n)\prod_{i=1}^n\dfrac{dx_i}{2i\pi}}.
\end{equation}

\section{Derivation of the entropic term in the matrix model effective action.}\label{App_entropic}
The Jacobian from the change of variable $d\phi_i$ to $D[\rho]$ can be obtained using the Faddeev-Poppov approach. We consider
\begin{equation}
1=\int{D[\rho_0]\prod_x\d\left(\rho_0(x)-\dfrac1N\sum_{i=1}^N\d(x-\phi_i)\right)},
\end{equation}
the product of delta functions can be represented with the help of a ghost field $\l(x)$ and gives
\begin{equation}
1=\int{D[\rho_0,\l]e^{i\int{dx\l(x)\rho_0(x)}}\prod_{i=1}^N e^{-\frac{i}{N}\l(\phi_i)}},
\end{equation}
where $2\pi$-factors were included in the measure $D[\l]$. It implies that
\begin{equation}
\int{\prod_{i=1}^Nd\phi_i}=\int{D[\rho_0,\l]e^{i\int{dx\l(x)\rho_0(x)}}\left(\int{e^{-\frac{i}{N}\l(\phi)}d\phi}\right)^N}.
\end{equation}
Or, introducing an effective action,
\begin{equation}
\int{\prod_{i=1}^Nd\phi_i}=\int{D[\rho_0,\l]e^{S[\rho_0,\l]}},\quad S[\rho_0,\l]=i\int{dx\l(x)\rho_0(x)}+N\log\left(\int{e^{-\frac{i}{N}\l(x)}dx}\right).
\end{equation}
At leading order, the integral over the ghost field can be evaluated as a saddle point. The equation of motion implies
\begin{equation}
\dfrac{\d S}{\d \l(x)}=0\implies \rho_0(x)=\g e^{-\frac{i}{N}\l(x)},\quad \text{with}\quad \g^{-1}=\int{e^{-\frac{i}{N}\l(x)}dx}.
\end{equation}
After replacing $\l(x)$ in the effective action, we notice that the $\g$-dependence drops since $\rho_0$ is normalized to one. We end up with the entropic term
\begin{equation}
\int{\prod_{i=1}^Nd\phi_i}=\int{D[\rho_0,\l]e^{-N\int{dx\rho_0(x)\log\rho_0(x)}}}.
\end{equation}

The subleading contribution is equal to the inverse of the square root of (minus) the Hessian determinant. The Hessian matrix evaluated at the saddle point gives
\begin{equation}
\dfrac{\d^2 S}{\d \l(x)\d\l(y)}=-\dfrac1N\rho_0(x)\left[\d(x-y)-\rho_0(y)\right].
\end{equation}
The factor of $1/N$ may be absorbed in the integration measure and will be discarded. The remaining determinant is a Fredholm determinant that can be computed exactly,
\begin{equation}
\det\left[-\dfrac{\d^2 S}{\d \l(x)\d\l(y)}\right]=e^{\int\log\rho_0(x)dx},
\end{equation}
In doing so, we removed a zero-mode associated to the unite norm of the density,
\begin{equation}
\det\left[\d(x-y)-\rho_0(y)\right]=1-\int{\rho_0(y)dy}.
\end{equation}
Thus, at the second order we have
\begin{equation}\label{sub_lead}
\int{\prod_{i=1}^Nd\phi_i}=\int{D[\rho_0]e^{-N\int{dx\rho_0(x)\log\rho_0(x)}-\frac12\int{dx\log\rho_0(x)}}},
\end{equation}
with factors $1/\sqrt{2\pi}N$ absorbed in the measure of integration (the factors $1/\sqrt{2\pi}$ cancel with the ones coming from the saddle point integration in \ref{ZC_1}).

\bibliographystyle{unsrt}

\end{document}